\PassOptionsToPackage{unicode}{hyperref}
\PassOptionsToPackage{hyphens}{url}
\PassOptionsToPackage{dvipsnames,svgnames,x11names}{xcolor}
\documentclass[
  12pt]{article}
\usepackage{xcolor}
\usepackage{amsmath,amssymb}
\usepackage{iftex}
\ifPDFTeX
  \usepackage[T1]{fontenc}
  \usepackage[utf8]{inputenc}
  \usepackage{textcomp} 
\else 
  \usepackage{unicode-math} 
  \defaultfontfeatures{Scale=MatchLowercase}
  \defaultfontfeatures[\rmfamily]{Ligatures=TeX,Scale=1}
\fi
\usepackage{lmodern}
\ifPDFTeX\else
\fi
\IfFileExists{upquote.sty}{\usepackage{upquote}}{}
\IfFileExists{microtype.sty}{
  \usepackage[]{microtype}
  \UseMicrotypeSet[protrusion]{basicmath} 
}{}
\makeatletter
\@ifundefined{KOMAClassName}{
  \IfFileExists{parskip.sty}{%
    \usepackage{parskip}
  }{
    \setlength{\parindent}{0pt}
    \setlength{\parskip}{6pt plus 2pt minus 1pt}}
}{
  \KOMAoptions{parskip=half}}
\makeatother
\makeatletter
\ifx\paragraph\undefined\else
  \let\oldparagraph\paragraph
  \renewcommand{\paragraph}{
    \@ifstar
      \xxxParagraphStar
      \xxxParagraphNoStar
  }
  \newcommand{\xxxParagraphStar}[1]{\oldparagraph*{#1}\mbox{}}
  \newcommand{\xxxParagraphNoStar}[1]{\oldparagraph{#1}\mbox{}}
\fi
\ifx\subparagraph\undefined\else
  \let\oldsubparagraph\subparagraph
  \renewcommand{\subparagraph}{
    \@ifstar
      \xxxSubParagraphStar
      \xxxSubParagraphNoStar
  }
  \newcommand{\xxxSubParagraphStar}[1]{\oldsubparagraph*{#1}\mbox{}}
  \newcommand{\xxxSubParagraphNoStar}[1]{\oldsubparagraph{#1}\mbox{}}
\fi
\makeatother

\usepackage{longtable,booktabs,array}
\usepackage{calc} 
\usepackage{etoolbox}
\makeatletter
\patchcmd\longtable{\par}{\if@noskipsec\mbox{}\fi\par}{}{}
\makeatother
\IfFileExists{footnotehyper.sty}{\usepackage{footnotehyper}}{\usepackage{footnote}}
\makesavenoteenv{longtable}
\usepackage{graphicx}
\makeatletter
\newsavebox\pandoc@box
\newcommand*\pandocbounded[1]{
  \sbox\pandoc@box{#1}%
  \Gscale@div\@tempa{\textheight}{\dimexpr\ht\pandoc@box+\dp\pandoc@box\relax}%
  \Gscale@div\@tempb{\linewidth}{\wd\pandoc@box}%
  \ifdim\@tempb\p@<\@tempa\p@\let\@tempa\@tempb\fi
  \ifdim\@tempa\p@<\p@\scalebox{\@tempa}{\usebox\pandoc@box}%
  \else\usebox{\pandoc@box}%
  \fi%
}
\def\fps@figure{htbp}
\makeatother

\providecommand{\tightlist}{%
  \setlength{\itemsep}{0pt}\setlength{\parskip}{0pt}}

\usepackage[]{natbib}
\usepackage{booktabs}
\usepackage{longtable}
\usepackage{array}
\usepackage{multirow}
\usepackage{wrapfig}
\usepackage{float}
\usepackage{colortbl}
\usepackage{pdflscape}
\usepackage{tabularx}
\usepackage{xltabular}
\usepackage{threeparttable}
\usepackage{threeparttablex}
\usepackage[normalem]{ulem}
\usepackage{makecell}
\usepackage{xcolor}

\usepackage{amsmath}
\usepackage{setspace}

\newcommand{\E}{\operatorname{E}}
\newcommand{\Prob}{\operatorname{P}}
\newcommand{\PP}{\mathbb{P}}
\newcommand{\var}{\mbox{Var}}

\newcommand{\expit}{\mbox{expit}}
\newcommand{\thetalim}{\theta^\ast}
\newcommand{\betalim}{\beta^\ast}
\newcommand{\PsiTildeY}{\Psi_{\tilde Y}}
\newcommand{\PsiDY}{\Psi_{\Delta Y}}
\newcommand{\PsiD}{\Psi_{\Delta}}
\newcommand{\PsiU}{\Psi_{U}}
\newcommand{\phiDY}{\phi_{\Delta Y}}
\newcommand{\phiTildeY}{\phi_{\tilde Y}}

\makeatletter
\@ifpackageloaded{caption}{}{\usepackage{caption}}
\AtBeginDocument{%
\ifdefined\contentsname
  \renewcommand*\contentsname{Table of contents}
\else
  \newcommand\contentsname{Table of contents}
\fi
\ifdefined\listfigurename
  \renewcommand*\listfigurename{List of Figures}
\else
  \newcommand\listfigurename{List of Figures}
\fi
\ifdefined\listtablename
  \renewcommand*\listtablename{List of Tables}
\else
  \newcommand\listtablename{List of Tables}
\fi
\ifdefined\figurename
  \renewcommand*\figurename{Figure}
\else
  \newcommand\figurename{Figure}
\fi
\ifdefined\tablename
  \renewcommand*\tablename{Table}
\else
  \newcommand\tablename{Table}
\fi
}
\@ifpackageloaded{float}{}{\usepackage{float}}
\floatstyle{ruled}
\@ifundefined{c@chapter}{\newfloat{codelisting}{h}{lop}}{\newfloat{codelisting}{h}{lop}[chapter]}
\floatname{codelisting}{Listing}

\makeatother
\makeatletter
\@ifpackageloaded{caption}{}{\usepackage{caption}}
\@ifpackageloaded{subcaption}{}{\usepackage{subcaption}}
\makeatother
\usepackage{bookmark}
\IfFileExists{xurl.sty}{\usepackage{xurl}}{} 
\hypersetup{
  pdftitle={One-step Outcome Imputation: An Alternative to Multiple Imputation},
  pdfauthor={Andreas Nordland; Klaus Kähler Holst; David Redek; Christian Pipper; Aske Thorn Iversen},
  pdfkeywords={Asymptotic variance, Influence function, Intercurrent
events, Missing data, Rubin's rules},
  colorlinks=true,
  linkcolor={blue},
  filecolor={Maroon},
  citecolor={Blue},
  urlcolor={Blue},
  pdfcreator={LaTeX via pandoc}}

\begin{document}

\def\spacingset#1{\renewcommand{\baselinestretch}%
{#1}\small\normalsize} \spacingset{1}


\date{November 3, 2026}
\title{\bf One-step Outcome Imputation: An Alternative to Multiple
Imputation}
\author{
Andreas Nordland\\
Novo Nordisk A/S\\
and\\Klaus Kähler Holst\\
Novo Nordisk A/S\\
and\\David Redek\\
Novo Nordisk A/S\\
and\\Christian Pipper\\
Novo Nordisk A/S\\
and\\Aske Thorn Iversen\\
Novo Nordisk A/S\\
}
\maketitle

\bigskip
\bigskip
\begin{abstract}
Missing outcomes in randomized controlled trials are often handled by
multiple imputation (MI). Rubin's rules are routinely used to estimate
standard errors but can fail to provide valid standard error estimates
for some commonly used procedures, such as reference-based imputation.
We propose a one-step alternative by explicitly targeting the treatment
effect implied by a given imputation model and constructing an efficient
one-step estimator for that treatment effect via its influence function.
Unlike Rubin's rules, this approach yields asymptotically valid
inference. Moreover, the proposed method circumvents the stochastic
component and computational burden of MI. We illustrate the approach
with examples spanning a range of imputation models, including
reference-based imputation, intercurrent-event-dependent imputation, and
repeated measures imputation. The one-step imputation estimation
procedure is implemented in the \texttt{targeted} R package.
\end{abstract}

\noindent%
\vfill

\newpage
\spacingset{1.9} 

\section{Introduction}\label{sec-intro}

Missing outcomes are common in randomized controlled trials (RCTs) and
are routinely handled by multiple imputation (MI), with uncertainty
quantified through Rubin's rules \citep{tan2021review}. However, the
validity of the analysis depends on alignment between the imputation
model and the complete-case estimator. In particular,
\citet{RobinsWang2000} describe the asymptotic bias of Rubin's rules for
both non-parametric and semiparametric analysis models. For the special
case of reference-based MI (also called controlled or placebo-based MI),
Rubin's rules lead to conservative inference and, consequently, to a
loss of power; see \citet{Bartlett02012023} for a recent review of
reference-based imputation in randomized clinical trials. We will also
present a clinically relevant example where Rubin's rules lead to
liberal inference.

We propose an alternative to MI in which we explicitly target the
treatment effect implied by a given imputation model. As we will
highlight, this is conceptually equivalent to performing infinite
multiple imputations based on that imputation model.

In our proposal the targeted treatment effect admits a natural
decomposition into the non-missing-outcome treatment effect and the
missing-outcome treatment effect under the imputation model. This
decomposition enables a formalized and transparent alignment between the
estimation procedure and the intercurrent-event strategy dictated by the
estimand \citep{ICHE9R1}. Specifically, the proposed procedure allows
for direct encoding of endpoint behavior following an intercurrent
event, in line with the chosen intercurrent-event strategy. Furthermore,
we use the decomposition to construct a semiparametric one-step
estimator that readily incorporates covariate adjustment via
randomization augmentation, leading to consistent and efficient
estimation in a randomized clinical trial \citep{FDAcovar}. The
corresponding influence function yields a closed-form expression for the
asymptotic variance of the one-step estimator.

Our one-step imputation procedure can be viewed as a generalization of
the conditional mean or regression imputation estimator
\citep{Rubin1987}. As noted by \citet{Rubin1987}, naive application of a
complete-case estimator to a single conditional-mean-imputed data set
will not provide valid inference: the variability due to the unknown
missing values is not correctly accounted for, leading to
downward-biased standard errors. Multiple imputation and other
resampling-based procedures such as the bootstrap and jackknife have
been proposed to address this issue
\citep{Rubin1987, Wolbers2022, Bartlett02012023}. The proposed one-step
estimation procedure solves this directly in a single step by explicitly
accounting for the uncertainty from estimating the imputation model
through its associated influence function, circumventing the need for
repeated imputation or resampling. We note that a similar procedure has
been proposed by \citet{KimRao2009} in the context of survey sampling.

In summary, our one-step procedure has several advantages compared to
MI:

\begin{enumerate}
\def\labelenumi{\arabic{enumi}.}
\tightlist
\item
  The procedure clearly states the target parameter without introducing
  additional structural assumptions. This allows the statistical
  properties of the estimator to be separated from discussion of its
  clinical interpretation. This is particularly relevant for
  reference-based imputation used in cases where outcomes are not deemed
  to be missing at random.
\item
  One-step imputation avoids the unnecessary stochastic component of MI.
  Estimation is less complex and computationally much faster, reducing
  the burden of the actual analysis as well as additional tipping-point
  analyses or simulation studies.
\item
  Unlike MI, the proposed one-step estimation procedure ensures
  asymptotically valid inference, and it achieves this without the
  potentially time-consuming resampling methods employed in the recent
  proposals by \citet{Bartlett02012023} and \citet{Wolbers2022}.
\end{enumerate}

The remainder of the paper is structured as follows. In
Section~\ref{sec-meth}, we present the methodology for a general
imputation model and provide a theoretical justification of the
proposal. In Section~\ref{sec-imp-examples}, we give concrete examples
of imputation models. In Section~\ref{sec-sim}, we report the results of
two simulation studies in which we investigate the empirical properties
of our proposal and benchmark it against MI. We conclude with a
discussion in Section~\ref{sec-dis}.

\section{Methods}\label{sec-meth}

\subsection{Setup \& Target Parameter}\label{sec-setup}

Let \(A\) denote a binary randomized treatment variable and let \(Y\)
denote the outcome of interest. If all outcomes are observed, our target
parameter is the usual average treatment effect \begin{align*}
\E[Y \mid A = 1] - \E[Y \mid A = 0].
\end{align*} However, if outcomes are missing, we only observe
\((\Delta, \Delta Y)\) for a non-missingness indicator \(\Delta\). Thus,
we cannot identify the average treatment effect without further
structural assumptions. Instead, we impute a replacement value \(U\) of
\(Y\) when missingness occurs, and the new target parameter becomes
\begin{align*}
\E[\tilde Y \mid A = 1] - \E[\tilde Y \mid A = 0],
\end{align*} where \begin{align*}
\tilde Y = \Delta Y + (1-\Delta) U.
\end{align*} We allow the replacement value \(U\) to depend on baseline
covariates \(X\), treatment \(A\), and post-randomization variables
\(Z\) through a known, sufficiently smooth functional parametrized in
\(\thetalim\): \begin{align*}
\tilde Y(\thetalim) = \Delta Y + (1-\Delta) U(X,A,Z;\thetalim).
\end{align*} Thus the target parameter becomes
\[\E[\tilde Y(\thetalim) \mid A = 1] - \E[\tilde Y(\thetalim) \mid A = 0]\]
We provide detailed clinically relevant examples of imputation models in
Section~\ref{sec-imp-examples}.

Usually, the imputation parameter \(\thetalim\) will have to be
estimated as well, which complicate the estimation procedure. In the
following section, we propose a one-step estimator and describe the
asymptotic distribution.

\subsection{Estimation \& Asymptotics}\label{sec-estimation}

We consider estimation of the target parameter
\(\PsiTildeY = \E[\tilde Y(\thetalim) \mid A = a]\), which we can
rewrite as \begin{align*}
\PsiTildeY &= \E[\tilde Y(\thetalim) \mid A = a]\\
&= \E\left[\Delta Y \mid A = a\right]
+ \Prob(\Delta = 0 \mid A = a)\, \E[U(X,A,Z;\thetalim) \mid \Delta = 0, A = a].
\end{align*} We aim to separately estimate each of the three following
sub-target parameters \begin{align*}
\PsiDY &= \E\left[\Delta Y \mid A = a\right],\\
\PsiD &= \Prob(\Delta = 0 \mid A = a),\\
\PsiU &= \E[U(X,A,Z;\thetalim) \mid \Delta = 0, A = a].
\end{align*} For each sub-target parameter, we calculate the influence
function of the corresponding estimator. We can then construct a plug-in
estimator for \(\PsiTildeY\) using the above formula and provide
inference via the delta method for influence functions
\citep{van1998asymptotic}.

The sub-target parameters \(\PsiDY\) and \(\PsiD\) are standard RCT
parameters for which consistent, regular asymptotically linear
estimators are readily available. We suggest applying a one-step
estimator adjusting for baseline covariates via a working outcome model
for robust and more efficient estimation; see \citet{KellyDukes2024} for
a recent review of covariate adjustment in an RCT setting. For \(n\) iid
observations and under weak assumptions for an estimated working outcome
model \(\widehat Q\), a robust one-step estimator for \(\PsiDY\) is
given by \begin{align*}
\widehat \PsiDY =
\frac{1}{n}\sum_{i = 1}^n \frac{I(A_i = a)}{\widehat g(a)} \left\{\Delta_i Y_i -
\widehat Q(X_i, A_i) \right\} + \widehat Q(X_i, a),
\end{align*} where \(\widehat g(a)\) is the empirical probability of
treatment \(a\). Furthermore, the associated influence function is given
by \begin{align*}
\sqrt{n}(\widehat \PsiDY - \PsiDY)
&= \frac{1}{\sqrt{n}}\sum_{i = 1}^n \left[\frac{I(A_i = a)}{g_0(a)} \left\{\Delta_i Y_i - Q^\ast(X_i, A_i) \right\} + Q^\ast(X_i, a) - \PsiDY \right] \\
&\quad + \frac{1}{\sqrt{n}}\sum_{i = 1}^n \frac{g_0(a) - I(A_i=a)}{g_0(a)} \E\left[\frac{I(A=a)}{g_0(a)}\left\{\Delta Y - Q^\ast(X,a) \right\}\right] \\
&\quad + o_{P_0}(1) \\
&= \frac{1}{\sqrt{n}}\sum_{i = 1}^n \phiDY(X_i, A_i, \Delta_iY_i; P^\ast) + o_{P_0}(1),
\end{align*} where \(g_0(a)>0\) is the probability of treatment
\(A = a\), \(Q^\ast\) is the limit of \(\widehat Q\), and
\(P^\ast = (g_0, Q^\ast, \E[Q^\ast], \PsiTildeY)\). When the outcome
model is correctly specified, i.e.,
\(Q^\ast(X, a) = \E[\Delta Y \mid X, A = a]\), the one-step estimator is
asymptotically efficient. The variance can be consistently estimated by
\(\frac{1}{n} \sum_{i = 1}^n \phiDY(X_i, A_i, \Delta_iY_i; \widehat P)^2\).
Note that we can construct one-step estimators that are (asymptotically)
equivalent to robust model-based treatment effect estimators such as
ANCOVA and G-computation \citep{Tsiatis2008, KellyDukes2024}. Similarly,
we can construct a one-step estimator for \(\PsiD\).

It remains to construct an estimator for \(\PsiU\). Regardless of the
application, we assume that a consistent, regular, and asymptotically
linear estimator \(\widehat{\theta}\) of \(\thetalim\) is available
based on the observed data. This will be the case for all the usual
parametric and semiparametric models, generalized linear models, mixed
models, survival models, and the like, fitted to observed data.
Specifically, we assume the linear decomposition \begin{align*}
\sqrt{n}(\widehat{\theta} - \thetalim) = n^{-1/2}\sum_{i=1}^n
\epsilon(X_i,A_i,Z_i,\Delta_i, \Delta_i Y_i; P_0) + o_{P_0}(1),
\end{align*} where \(\epsilon\) is the influence function associated
with \(\widehat \theta\). For maximum likelihood type estimators, the
influence function is proportional to the score, i.e., the derivative of
the log-likelihood. Note that if \(\widehat \theta\) is fitted on a
subset of the observed data, only entries of \(\epsilon\) associated
with this subset will be non-zero. Detailed examples, including
subset-based regression models and intercurrent-event-dependent
imputation models, are given in Section~\ref{sec-imp-examples}.

By the delta method for influence functions, the sampling variability of
\(U(x, a, z; \widehat \theta)\) inherited from \(\widehat \theta\)
satisfies \begin{align}
&\sqrt{n}[U(x,a,z; \widehat \theta) - U(x,a,z; \thetalim)] \nonumber \\ =
&\frac{1}{\sqrt{n}}\sum_{i=1}^n \nabla_\theta U(x,a,z; \thetalim)\, \epsilon(X_i,A_i,Z_i,\Delta_i,
\Delta_i Y_i; P_0) + o_{P_0}(1), \label{eq:UthetaIF}
\end{align} where \(\nabla_{\theta}\) is the partial derivative operator
with respect to \(\theta = (\theta_1, \ldots, \theta_k)^\top\),
\begin{align*}
\nabla_{\theta} U(\theta) = \left(\frac{\partial}{\partial \theta_1} U(\theta)
\quad \cdots \quad \frac{\partial}{\partial \theta_k} U(\theta)\right).
\end{align*} A plug-in estimator for
\(\PsiU = \E[U(X, A, Z; \thetalim) \mid \Delta = 0, A = a]\) is given by
\[
\widehat \PsiU = \frac{1}{n} \sum_{i=1}^n \frac{I(A_i=a)}{\widehat
g(a)}\frac{1-\Delta_i}{1 - \widehat
S(A_i)} U(X_i, A_i, Z_i; \widehat \theta),
\] where \(\widehat g(a)\) is the empirical estimate of the probability
of receiving treatment \(a\), and \(\widehat S(a)\) is the empirical
estimate of the conditional probability of not being missing in
treatment stratum \(a\), which we assume to be positive, i.e.,
\(S_0(a)>0\). As shown in Supplementary Material~\ref{sec-supp-asymp},
it holds that \begin{align*}
\sqrt{n}&(\widehat \PsiU - \PsiU) \\
=& n^{-1/2} \sum_{i = 1}^n \frac{I(A_i=a)}{g_0(a)}\frac{1-\Delta_i}{1-S_0(a)}\left\{U(X_i, A_i, Z_i; \thetalim)
- \PsiU\right\}\\
&+ n^{-1/2} \sum_{i = 1}^n \PP\left(\frac{I(A=a)}{g_0(a)}\frac{1-\Delta}{1 -
S_0(a)}\nabla_{\theta} U(X,A,Z; \thetalim)\right) \epsilon(X_i, A_i, Z_i, \Delta_i,
\Delta_i Y_i; P_0)\\
&+ o_{P_0}(1).
\end{align*} As with the other sub-target parameters, treatment
randomization makes it possible to gain efficiency via covariate
adjustment (augmentation). This is outlined in
Section~\ref{sec-supp-asymp-2}.

The component-wise estimator described above is instructive in that it
isolates the treatment effect implied by the imputation model. However,
it is also possible to construct a direct one-step imputation estimator,
as outlined in Section~\ref{sec-supp-direct}. An advantage of this
approach is that it requires only a single working outcome model and
that its specification is almost identical to that of the standard
multiple imputation approach. We discuss the direct one-step imputation
estimator further in Section~\ref{sec-dis}.

\subsection{Link to Multiple Imputation}\label{sec-link-mi}

We formulate the conditional multiple-imputation procedure following
\citet{RobinsWang2000}:

\begin{enumerate}
\def\labelenumi{\arabic{enumi}.}
\tightlist
\item
  Let \(f(y|X,A,Z, \Delta = 1; \beta)\) denote the observed conditional
  outcome likelihood parametrized in \(\beta\). Let \(\widehat \beta\)
  be a regular and asymptotically linear estimator with limit
  \(\betalim\). Usually, we let \(\widehat \beta\) be the MLE.
\item
  Create \(m\) complete data sets by sampling missing outcomes from
  \(f(y|X,A,Z, \Delta = 1; \widehat \beta)\).
\item
  Apply the analysis estimator to each of the \(m\) completed data sets.
\item
  Let the multiple imputation estimate be the empirical mean over the
  \(m\) complete data analysis estimates.
\end{enumerate}

The imputation function \(U(X, A, Z; \theta)\) described in this paper
is closely related to \(f(y|X,A,Z,\Delta = 1; \beta)\) in the sense that
we can express \(U(X, A, Z; \theta)\) as a function of
\(f(y|X,A,Z,\Delta = 1; \beta)\). For example, for reference-based
imputation we let \[
U(X,Z;\theta(\beta)) = \int y\, f(y|X, A=0, Z; \beta)\, dy.
\] In this article we focus on a one-step estimator as the complete-data
analysis estimator; see Section~\ref{sec-estimation}. Let \(U_{ij}\)
denote the \(j\)th imputation for observation \(i\) in Step 2 of the
multiple imputation procedure. Then the multiple-imputation estimator of
the expected imputed outcome in treatment arm \(a\) is given by
\begin{align}
&\frac{1}{n}\sum_{i = 1}^n  \frac{I(A_i = a)}{\widehat g(a)} \left\{\Delta_i
Y_i + (1-\Delta_i) \left(\frac{1}{m}\sum_{j = 1}^m U_{ij} \right) - \widehat
Q(X_i, A_i) \right\} + \widehat
Q(X_i, a) \nonumber \\
=& \frac{1}{n}\sum_{i = 1}^n  \frac{I(A_i = a)}{\widehat g(a)} \left\{\Delta_i
Y_i - \widehat
Q(X_i, A_i) \right\} + \widehat
Q(X_i, a) \label{eq:mi_est_term1} \\
+& \frac{1}{n}\sum_{i = 1}^n  \frac{I(A_i = a)}{\widehat g(a)} (1-\Delta_i)
\left(\frac{1}{m}\sum_{j = 1}^m U_{ij} \right). \label{eq:mi_est_term2}
\end{align} Term \eqref{eq:mi_est_term1} converges in probability to
\(\E[\Delta Y \mid A = a]\) under weak assumptions on the outcome model
\(\widehat Q\). For term \eqref{eq:mi_est_term2}, the law of large
numbers gives, as \(m \to \infty\), \begin{align*}
\frac{1}{m}\sum_{j = 1}^m U_{ij} \overset{a.s.}{\to} \int y\, f(y|X_i, A_i, Z_i, \Delta_i
= 1; \widehat \beta)\, dy.
\end{align*}

As outlined in Section~\ref{sec-supp-mi-target}, for
\(m, n \to \infty\), we get that \begin{align*}
 \frac{1}{n} \sum_{i = 1}^n (1-\Delta_i) \left(\frac{1}{m}\sum_{j = 1}^m
 U_{ij}\right)
 \overset{P}{\to} &\E\left[(1-\Delta)\int y\, f(y|X, A, Z, \Delta
= 1; \betalim)\, dy\right]\\
= &\E\left[(1-\Delta)U(X,A,Z;\thetalim)\right],
\end{align*} where \begin{align*}
U(X,A,Z;\thetalim) = \int y\, f(y|X, A, Z; \betalim)\, dy.
\end{align*} Similar results are shown in \citet{Wolbers2022} for the
ANCOVA model and in \citet{Yang2016} for a Bayesian multiple-imputation
approach.

\section{Imputation Model Examples}\label{sec-imp-examples}

The preceding development of the imputation model and its asymptotic
theory via the influence function has been deliberately general. This
section makes the framework concrete by working through four examples of
practical relevance for clinical trials. We first show how to compute
the influence function for a reference-based imputation model, using a
logistic model as the working model
(Section~\ref{sec-subset-M-estimator}). We then describe how the
framework accommodates intercurrent events under different strategies,
in line with the ICH E9(R1) addendum: in Section~\ref{sec-ICE}, we
consider a treatment-policy strategy, working through a
retrieved-dropout imputation model, and in Section~\ref{sec-ICE2}, we
consider a hypothetical strategy. Finally, in longitudinal trials, the
outcome may be measured repeatedly over time. In
Section~\ref{sec-repeated-measures-imputation} we sketch how to
construct an imputation model based on a mixed model for repeated
measures.

We highlight that the concepts and techniques introduced in the examples
can be combined to formulate a wide range of clinically relevant
imputation models.

\subsection{Reference-based imputation}\label{sec-subset-M-estimator}

As described by \citet{carpenter2013analysis}, the intuition behind
reference-based imputation is to construct an imputation model that
approximates the observed-data conditional outcome in the reference
(control/placebo) arm, i.e., \[
U(X, Z; \thetalim) \approx \E[Y \mid X, A=0, Z, \Delta = 1].
\] In a superiority-trial setting, this model can be used as a
worst-case scenario that shrinks the treatment effect. Typically, we
formulate a parametric or semiparametric imputation working model, e.g.,
a generalized linear model or a mixed model, which allows us to estimate
\(\thetalim\) and the associated influence function via maximum
likelihood.

Recall that maximum likelihood estimators are a special case of
M-estimators, i.e., estimators of a \(k\)-dimensional parameter
\(\thetalim\) defined by the \(k\)-dimensional score equation
\begin{align*}
\E[m(X, A, Z, \Delta, \Delta Y; \thetalim)] = 0^{k\times 1}.
\end{align*} Under suitable regularity conditions, the influence
function for the estimator \(\widehat\theta\) defined as the solution to
\begin{align*}
\sum_{i = 1}^n m(X_i, A_i, Z_i, \Delta_i, \Delta_i Y_i;\widehat \theta) = 0^{k\times 1}
\end{align*} is given by \begin{align*}
\E\left[- \nabla_{\theta} m(X, A, Z, \Delta, \Delta Y; \thetalim)
\right]^{-1} m(X, A, Z, \Delta, \Delta Y;\thetalim).
\end{align*}

Thus, for a logistic model with covariate vector \(W\) (including an
intercept) and outcome \(Y\), the score is
\(m(W,Y; \theta) = W(Y - p(W; \theta))\) and the influence function is
\begin{align*}
\E\left[p(W; \theta)\{1 - p(W; \theta)\} W W^\top
\right]^{-1} W (Y - p(W; \theta)),
\end{align*} where \(p(W; \theta) = \expit(W^\top \, \theta)\). See
Section~\ref{sec-supp-mi-target} for more details on the exponential
family in general.

In reference-based imputation, the model is fitted on a subset of the
observed data, specifically the control arm completers
\(\{A = 0, \Delta =
1\}\). On this subset, the score reduces to a function of \((X, Z, Y)\)
and the maximum likelihood estimator solves \begin{align*}
\sum_{i = 1}^n R_i\, m(X_i, Z_i, Y_i;\widehat \theta) = 0^{k\times 1},
\end{align*} where \(R = I(A = 0)\,I(\Delta = 1)\). The associated
influence function is given by \begin{align*}
\epsilon(X, A, Z, \Delta, \Delta Y; P_0) = \frac{R}{\Prob(R=1)} \E \left[- \nabla_{\theta} m(X,Z, Y; \thetalim)
\,\Big|\, R = 1 \right]^{-1} m(X,Z,Y;\thetalim).
\end{align*}

\subsection{Retrieved dropout imputation}\label{sec-ICE}

The inclusion of the post-randomization variable \(Z\) in the imputation
model \(U(X, A, Z; \thetalim)\) naturally allows for different
imputation strategies depending on the occurrence of intercurrent events
(ICEs). Retrieved-dropout imputation is commonly used for estimands
where the ICE is treatment discontinuation handled under a
treatment-policy strategy; see, e.g., \citet{McEvoy02012016}.

Let \(\Gamma = \Gamma(Z)\) be an indicator for the ICE. To illustrate
the form of the imputation model in this setting, we make the
stratification via \(\Gamma\) explicit by writing the imputation
function as \begin{align*}
U(X, A, Z; \theta) = \Gamma \cdot U_{\Gamma}(X, A, Z; \theta_{\Gamma})
+ (1-\Gamma) \cdot U_{1-\Gamma}(X, A, Z; \theta_{1-\Gamma}),
\end{align*} where \(U_{\Gamma}\) specifies the imputation strategy for
subjects who experience the ICE and \(U_{1-\Gamma}\) specifies the
strategy for subjects who do not.

Usually the two imputation models are fitted on disjoint subsets defined
by the occurrence of the ICE. Specifically, for treatment
discontinuation we distinguish between retrieved dropouts
\(\{\Gamma = 1, \Delta
= 1\}\) and observed-on-drug \(\{\Gamma = 0, \Delta
= 1\}\). Then \begin{align*}
U_{\Gamma}(X, A, Z; \thetalim_{\Gamma}) &\approx \E[Y \mid X, A, Z, \Gamma = 1,
\Delta = 1],\\
U_{1-\Gamma}(X, A, Z; \thetalim_{1-\Gamma}) &\approx \E[Y \mid X, A, Z, \Gamma =
0, \Delta = 1].
\end{align*} This formulation makes the assumptions for each imputation
model explicit and separable, in line with the structured approach
advocated by \citet{ICHE9R1}.

As in Section~\ref{sec-subset-M-estimator}, we can calculate the
influence functions for \(\widehat\theta_{\Gamma}\) and
\(\widehat\theta_{1-\Gamma}\) separately, and an application of the
delta method gives \begin{align*}
\sqrt{n}&[U(x,a,z; \widehat \theta) - U(x,a,z; \thetalim)] \\
=& \frac{1}{\sqrt{n}}\sum_{i=1}^n \Gamma(z) \, \nabla_{\theta_{\Gamma}} U_{\Gamma}(x,a,z;
\thetalim_{\Gamma}) \epsilon_{\Gamma}(X_i, A_i, Z_i, \Delta_i, \Delta_i Y_i; P_0) \\
+&\frac{1}{\sqrt{n}}\sum_{i=1}^n (1-\Gamma(z))\, \nabla_{\theta_{1-\Gamma}} U_{1-\Gamma}(x,a,z;
\thetalim_{1-\Gamma}) \epsilon_{1-\Gamma}(X_i, A_i, Z_i, \Delta_i, \Delta_i Y_i; P_0)\\
+& o_{P_0}(1).
\end{align*}

\subsection{A hypothetical strategy}\label{sec-ICE2}

In this example, we consider a hypothetical strategy for an ICE as
defined in \citet{ICHE9R1}. It is customary not to utilize outcome
measures after ICEs under a hypothetical strategy; see, e.g.,
\citet{Parra03042023}. In the following, we adopt this approach.

As a simple case, let \(\Gamma = \Gamma(Z)\) denote the indicator for
the occurrence of an intercurrent event. Following \citet{ICHE9R1}, we
make a precise distinction between two fundamentally different sources
of missingness:

\begin{itemize}
\tightlist
\item
  \(\{\Delta = 0, \Gamma = 0\}\): \textbf{Missing data} - no ICE
  occurred, but the outcome is missing, for example due to loss to
  follow-up.
\item
  \(\{\Gamma = 1\}\): \textbf{Structurally missing} - the outcome is not
  meaningful for the treatment effect as a direct consequence of the ICE
  strategy, regardless of whether it was collected.
\end{itemize}

To unify both sources of missingness, we define the effective
non-missingness indicator \begin{align*}
\Delta(\Gamma) = \Delta \cdot (1 - \Gamma),
\end{align*} which enforces that whenever an ICE occurs the outcome is
treated as missing. The imputed outcome now becomes \begin{align*}
\tilde{Y} = \Delta(\Gamma) Y + (1 - \Delta(\Gamma)) U(X, A, Z, \Delta; \thetalim),
\end{align*} which has exactly the same form as the imputed outcome in
Section~\ref{sec-setup}.

More generally, in a trial with multiple ICEs and possibly also several
imputation strategies for missing data, all relevant post-randomization
information, including ICE indicators and their timing, can be collected
into the post-randomization variable \(Z\). The effective
non-missingness indicator is then defined appropriately to enforce
missingness for all structurally missing outcomes, and the imputation
function \(U(X, A, Z, \Delta; \thetalim)\) encodes the full set of
imputation strategies.

\subsection{Repeated measures
imputation}\label{sec-repeated-measures-imputation}

Mixed models for repeated measures are a common choice of imputation
model in longitudinal studies with a repeatedly measured continuous
outcome. For illustrative purposes, we consider a simple two-time-point
setup, but this can be generalized to any multi-time-point mean and
covariance structure.

Assume that we observe baseline covariates \(X\), treatment \(A\), and
outcomes \(Y_1\) and \(Y_2\). We are interested in the imputed average
treatment effect for the outcome \(Y = Y_2\). Both \(Y_1\) and \(Y_2\)
can be missing. Let \(\Delta_1\) and \(\Delta_2\) denote the associated
non-missingness indicators. Using the notation introduced in
Section~\ref{sec-setup}, we let \(Z = (\Delta_1, \Delta_1 Y_1)\),
\(\Delta = \Delta_2\) and \(Y = Y_2\).

We formulate a linear model for \((Y_1, Y_2)\) given \(X\) and \(A\) as
\begin{align*}
\begin{pmatrix}
Y_1\\
Y_2
\end{pmatrix} \sim \mathcal{N}\Big(\mu(X, A; \beta), \Sigma \Big),
\end{align*} where
\(\mu(X, A; \beta) = (\mu_1(X,A;\beta), \mu_2(X,A;\beta))^\top\) is a
known function parameterized in \(\beta\) and \(\Sigma\) is the
covariance matrix with entries
\((\sigma_1^2, \sigma_2^2, \sigma_{12})\). Under a missing-at-random
assumption, the likelihood contribution for a single observation is
proportional to \begin{align*}
&f(\Delta_1 Y_1, \Delta_2 Y_2 \mid X, A, \Delta_1, \Delta_2; \theta) \\
&= I(\Delta_1 = 1, \Delta_2 = 0)
\left\{\sigma_1^{-1}\exp\!\Big(-2^{-1}[Y_1 -
\mu_1(X,A;\beta)]^2\sigma_1^{-2}\Big)\right\}\\
&\quad + I(\Delta_1 = 0, \Delta_2 = 1)
\left\{\sigma_2^{-1}\exp\!\Big(-2^{-1}[Y_2 -
\mu_2(X,A;\beta)]^2\sigma_2^{-2}\Big)\right\}\\
&\quad + I(\Delta_1 = 1, \Delta_2 = 1)
\left\{\det(\Sigma)^{-1/2}\exp\!\Big(-2^{-1}(Y -
\mu(X,A;\beta))^\top \Sigma^{-1}(Y -
\mu(X,A;\beta))\Big)\right\},
\end{align*} where
\(\theta = (\beta, \sigma_1^2, \sigma_2^2, \sigma_{12})^\top\). Thus, we
can construct a maximum likelihood estimator \(\widehat \theta\) with
limit \(\thetalim\). Furthermore, the influence function
\(\epsilon(X,A, Z,\Delta,
\Delta Y; P_0)\) can be calculated via the associated score function
\citep{tsiatis2006semiparametric}.

Finally, the imputation function is given by the conditional-mean
predictions implied by the fitted linear model \begin{align*}
U(X,A,Z; \theta) =
\begin{cases}
\mu_2(X,A; \beta) & \text{if } \Delta_1 = 0, \\
\mu_2(X,A; \beta) + \sigma_{12}\, \sigma_1^{-2}\{Y_1 - \mu_1(X,A;\beta)\}
& \text{if } \Delta_1 = 1.
\end{cases}
\end{align*}

\section{Simulations}\label{sec-sim}

We investigate the empirical properties of the proposed one-step
imputation estimator through two simulation studies based on the
examples in Section~\ref{sec-imp-examples}. In each case, we benchmark
against the associated MI estimator. In the reference-based imputation
simulation described in Section~\ref{sec-simref}, we confirm that the MI
procedure is conservative. By contrast, in the retrieved-dropout
imputation simulation described in Section~\ref{sec-simdrop}, we find
the MI procedure to be liberal. In both cases, our one-step imputation
estimator provides valid inference.

The one-step imputation estimator is implemented in the
\texttt{targeted} R package; see \citet{targeted090}. The MI estimation
procedure is described in Section~\ref{sec-link-mi}, and the standard
error estimates are pooled using Rubin's rules as implemented in the
\texttt{mice} R package; see \citet{mice2011}.

\subsection{Reference-based imputation}\label{sec-simref}

We consider the following data-generating mechanism.

\begin{itemize}
\tightlist
\item
  Treatment: \(A \sim \text{Bernoulli}(0.5)\)
\item
  Baseline covariates: \begin{align*}
  X_1 & \sim \text{Bernoulli}\big(\expit\{0.25\}\big)\\
  X_2 \mid X_1 & \sim \mathcal{N}(X_1, 1)
  \end{align*}
\item
  Non-missingness: \begin{align*}
  \Delta \mid A, X_1, X_2 \sim \text{Bernoulli}\big(\expit\big\{&1.75 + 0.3A + 0.1X_1 - 0.3X_2 + 0.8AX_1\\
  &+ 0.2AX_2 + 0.4X_1X_2 - 1.8AX_1X_2\big\}\big)
  \end{align*}
\item
  Outcome: \begin{align*}
  Y \mid A, X_1, X_2, \Delta = 1 \sim \text{Bernoulli}\big(\expit\big\{&-0.6 + 1.1A - 0.3X_2 + 0.6AX_2\\
  &+ 0.5X_1X_2 + 0.1AX_1X_2\big\}\big)
  \end{align*}
\end{itemize}

We note that we do not include any post-randomization variables. The
proportion of missing outcomes in this setting is approximately 14\% in
the control arm (\(A = 0\)) and 20\% in the treatment arm (\(A = 1\)).

We define the target parameter as the imputed average treatment effect
associated with the logistic model \begin{align*}
\E[Y \mid A = 0, X_1, X_2, \Delta = 1] \approx U(X_1, X_2; \thetalim) = \mbox{expit}\left(\thetalim_1 +
\thetalim_2 X_1 + \thetalim_3 X_2 + \thetalim_4 X_1 X_2\right).
\end{align*} Note that the model is fitted on the non-missing reference
subset \(\{A = 0, \Delta = 1\}\). In this particular case, the true
limiting parameter vector is \(\thetalim = (-0.6, 0, -0.3, 0.5)^\top\).

For the one-step imputation estimator, the sub-target parameters
\(\E[\Delta Y \mid A = a]\) and \(\Prob(\Delta = 0 \mid A = a)\) are
estimated via the one-step estimator described in
Section~\ref{sec-estimation}, using a logistic outcome model fitted by
maximum likelihood with all interactions between \(A\), \(X_1\), and
\(X_2\). The imputation sub-target parameter
\(\E[U(X, A; \thetalim) \mid A = a, \Delta = 0]\) is estimated via the
plug-in estimator described in Section~\ref{sec-estimation}.

We also apply the MI procedure using a one-step estimator as the
analysis estimator applied to each imputed data set. We therefore expect
the one-step imputation and MI estimators to have almost identical
distributions. However, we expect the MI standard error estimates to be
positively biased, leading to conservative inference. The MI estimator
is based on \(100\) imputation repetitions.

We approximate the target parameter using Monte Carlo sampling (\(10^7\)
replications) based on evaluations of the known outcome, missingness
probability, and imputation models.

The simulation results are summarized in Table~\ref{tbl-simref}. We
confirm that the MI and one-step imputation estimators are both unbiased
and have similar empirical means and standard deviations. However, the
estimated MI standard errors are on average \(20\%\) higher than the
standard deviation, resulting in a coverage of \(98.4\%\) for a nominal
\(95\%\) Wald confidence interval. In comparison, the mean standard
error of the one-step imputation estimator match the empirical standard
deviation closely, achieving the nominal coverage.

\begin{table}

\caption{\label{tbl-simref}Simulation results for reference-based
imputation across \(R = 10{,}000\) Monte Carlo replications of sample
size \(n = 2{,}000\).}

\centering{

\centering
\begin{tabular}{lrr}
\toprule
Metric & MI & One-step\\
\midrule
Empirical mean & 0.24441 & 0.24437\\
Standard deviation (SD) & 0.01938 & 0.01941\\
Empirical mean standard error (SE) & 0.02332 & 0.01920\\
SE/SD & 1.20360 & 0.98895\\
Target parameter & 0.24432 & 0.24432\\
Bias & 0.00009 & 0.00005\\
Coverage (95\% Wald confidence interval) & 0.98390 & 0.94770\\
\bottomrule
\end{tabular}

}

\end{table}%

\subsection{Retrieved dropout imputation}\label{sec-simdrop}

In this simulation example we consider a more complex setting with a
composite outcome for handling terminal events and a retrieved dropout
strategy for handling missing outcomes. The setup is inspired by the
recently conducted FLOW trial \citet{perkovic2024effects}. Specifically,
after randomization we collect a treatment discontinuation indicator
\(Z_1\) and a non-terminal event indicator \(Z_2\). If a terminal event
occurs (\(Z_2 = 0\)), we assign a fixed outcome value. Otherwise, we
measure the outcome of interest, which may be missing.

The data-generating mechanism is as follows.

\begin{itemize}
\tightlist
\item
  Treatment: \(A \sim \text{Bernoulli}(0.5)\).
\item
  Baseline covariates: \begin{align*}
  X_1 & \sim \text{Bernoulli}(\pi_{X_1})\\
  X_2 \mid X_1 & \sim \mathcal{N}(\mu_{X_2}(X_1), 1)\\
  X_3 \mid X_1, X_2 & \sim \mathcal{N}(\mu_{X_3}(X_1, X_2), 1).
  \end{align*}
\item
  Treatment discontinuation: \begin{align*}
  Z_1 \mid A, X_1, X_2, X_3 \sim \text{Bernoulli}(\pi_{Z_1}\{A, X_1, X_2, X_3\}),
  \end{align*} where \(\pi_{Z_1}\) is an \(8\)-term logistic function in
  \(A, X_1, X_2, X_3\) including all interactions of \(A\) with
  \((X_1, X_2, X_3)\).
\item
  Non-terminal event: \begin{align*}
  Z_2 \mid A, X_1, X_2, X_3, Z_1 \sim \text{Bernoulli}(\pi_{Z_2}\{A, X_1, X_2, X_3, Z_1\}),
  \end{align*} where \(\pi_{Z_2}\) is a \(16\)-term logistic function
  including all interactions of \(A\) and \(Z_1\) with
  \((X_1, X_2, X_3)\) and the three-way interactions \(AZ_1 X_j\) for
  \(j = 1, 2, 3\).
\item
  Non-missingness: \begin{align*}
  \Delta \mid A, X_1, X_2, X_3, Z_1 \sim \text{Bernoulli}(1 - \pi_{\text{mis}}\{A, X_1, X_2, X_3, Z_1\}),
  \end{align*} where \(\pi_{\text{mis}}\) has the same \(16\)-term
  logistic function structure as \(\pi_{Z_2}\).
\item
  Outcome: \begin{align*}
  Y \mid A, X_1, X_2, X_3, Z_1, Z_2, \Delta = 1 \sim
  \begin{cases}
  \mathcal{N}(\mu_Y\{A, X_1, X_2, X_3, Z_1\}, 1) & \text{if } Z_2 = 1, \\
  -1.6 & \text{if } Z_2 = 0,
  \end{cases}
  \end{align*} where \(\mu_Y\) is a linear function with \(16\) terms
  with the same interactions as \(\pi_{Z_2}\) and \(\pi_{\text{mis}}\).
  Note that we only observe \(\Delta Y\).
\end{itemize}

The full parametrization of
\(\pi_{X_1}, \mu_{X_2}, \mu_{X_3}, \pi_{Z_1},
\pi_{Z_2}, \pi_{\text{mis}}\), and \(\mu_Y\) is given in
Section~\ref{sec-supp-siminter-dgp}. Under this specification, treatment
discontinuation occurs in approximately \(18\%\) of subjects and
terminal events in approximately \(14\%\). Among subjects with no
terminal event, \(9\%\) of the outcomes are missing.

As in the previous example, we define the target parameter as the
imputed average treatment effect associated with a retrieved-dropout
imputation model defined as the limit of the linear regression
\begin{align*}
\E[Y \mid A, X_1, X_2, X_3, Z_1, Z_2 = 1, \Delta = 1] \approx U(X,A,Z;\thetalim)
= (1, A, X_1, X_2, X_3, Z_1, AZ_1)^\top\, \thetalim.
\end{align*} Note that the imputation model is fitted on the non-missing
non-terminal subset \(\{\Delta = 1, Z_2 = 1\}\).

For the one-step imputation estimator, the sub-target parameters
\(\E[\Delta Y \mid A = a]\) and \(\Prob(\Delta = 0 \mid A = a)\) are
estimated via the one-step estimator described in
Section~\ref{sec-estimation}, using a linear outcome model and a
logistic regression outcome model, respectively, both fitted by maximum
likelihood with main effects for \(A\), \(X_1\), \(X_2\), and \(X_3\).
The imputation sub-target parameter
\(\E[U(X, A, Z; \thetalim) \mid A = a, \Delta = 0]\) is estimated via
the plug-in estimator described in Section~\ref{sec-estimation}.

The MI procedure uses a one-step analysis estimator applied to each
imputed data set based on the same linear regression outcome working
model. The procedure is based on \(100\) imputation repetitions.

We approximate the target parameter using nested Monte Carlo sampling.
Specifically, we sample \(10^6\) observations from the data generation
mechanism, fit the imputation model, and compute the plug-in estimator
for the imputed average treatment effect. This is repeated \(500\)
times, and the average of the plug-in estimates is used as the target
parameter approximation.

The simulation results are summarized in Table~\ref{tbl-siminter}. As
expected, the MI and one-step estimators are both unbiased and have
similar empirical distributions. The mean standard error of the one-step
estimator is almost identical to the standard deviation, and the
associated Wald-type confidence intervals have a coverage of \(94.8\%\).
In comparison, the mean standard error of the MI estimator based on
Rubin's rules is \(3.6\%\) lower than the standard deviation, and the
coverage of the Wald-type confidence intervals is \(94.1\%\). Thus, in
this setting, inference based on Rubin's rules is liberal.

\begin{table}

\caption{\label{tbl-siminter}Simulation results for retrieved dropout
imputation across \(R = 10{,}000\) Monte Carlo replications of sample
size \(n = 4{,}000\). Columns compare the multiple-imputation (MI)
estimator with the proposed one-step imputation estimator for the
imputed average treatment effect.}

\centering{

\centering
\begin{tabular}{lrr}
\toprule
Metric & MI & One-step\\
\midrule
Empirical mean & 0.20834 & 0.20832\\
Standard deviation (SD) & 0.03722 & 0.03725\\
Empirical mean standard error (SE) & 0.03588 & 0.03707\\
SE/SD & 0.96385 & 0.99502\\
Target parameter & 0.20808 & 0.20808\\
Bias & 0.00026 & 0.00024\\
Coverage (0.95 Wald confidence interval) & 0.94060 & 0.94820\\
\bottomrule
\end{tabular}

}

\end{table}%

\section{Discussion}\label{sec-dis}

One-step imputation is an attractive alternative to multiple imputation
in randomized clinical trials. It avoids the stochastic and
computationally costly aspects of multiple imputation while always
providing valid asymptotic inference. Furthermore, by explicitly
formulating the treatment effect measure implied by a given imputation
model, the framework disentangles estimation properties from structural
or causal assumptions.

Throughout the article we have intentionally avoided formulating any
structural or causal assumptions justifying the choice of the considered
imputation models. Similarly, we do not claim that the imputation model
equals a given conditional expectation. We note however that in any
application the imputation model should be justified from a causal
perspective along with a sensitivity analysis of the possible
misspecification of the imputation model.

In the specific case where a missing-at-random assumption can be
justified, e.g., \(Y \perp \Delta \mid X, A, Z\), we advocate for the
use of (efficient) doubly robust estimators for the treatment effect. A
doubly robust one-step estimator for the causal target parameter
\(\E[Y \mid A=a]\) is then given by \begin{align*}
&\frac{1}{n} \sum_{i=1}^n \frac{\Delta_i}{\widehat
S(X_i,A_i,Z_i)}\frac{I(A_i=a)}{\widehat g(a)} Y_i + \frac{\widehat g(a) -
I(A_i=a)}{\widehat g(a)}\widehat Q_1(X_i,a) \\
+ &\frac{1}{n} \sum_{i=1}^n \frac{I(A_i=a)}{\widehat g(a)}
\frac{\widehat S(X_i,A_i,Z_i)-\Delta_i}{\widehat S(X_i,A_i,Z_i)} \widehat
Q_2(X_i,A_i, Z_i),
\end{align*} where \(\widehat S\) is a model for
\(\E[\Delta \mid X, A, Z]\), \(\widehat Q_2\) is a model for
\(\E[Y \mid X, A, Z] = \E[Y \mid X, A, Z, \Delta = 1]\), and \(\widehat
Q_1\) is a model for
\(\E[Y|X, A] = \E[\E[Y|X, A, Z, \Delta = 1]|X, A]\). If the missingness
model \(\widehat S\) or the outcome model \(\widehat Q_2\), is correctly
specified, the above estimator is a consistent estimator of the causal
target parameter. However, the full advantage of the doubly robust
estimation procedure is realized only if we allow for flexible modeling
of the nuisance models. Future work will focus on combining conservative
imputation strategies for certain intercurrent events with
missing-at-random assumptions for non-structural missing outcomes.

A further direction for future work is the practical handling of more
complicated missingness patterns in baseline and post-randomization
variables used in the imputation model. This includes a general
implementation of mixed models for repeated measures as sketched in
Section~\ref{sec-repeated-measures-imputation}.

As outlined in Section~\ref{sec-estimation} and
Section~\ref{sec-supp-direct}, it is possible to construct a direct
one-step imputation estimator with a simpler specification than the
sub-target parameter approach. It is unclear which estimator has better
operating characteristics in terms of precision. The results of
\citet{cohen2023harm} and \citet{bannick2025general} indicate that a
direct estimation approach may provide stronger efficiency gain
guarantees. However, their results are not directly applicable due to
the additional influence function term related to the imputation
parameters. Future work will investigate efficiency calibration under
outcome imputation.

A final outstanding issue is that inference deduced from one-step
imputation estimation, like all Wald-type estimators, is sensitive to
small-sample issues. Future work will investigate existing small-sample
corrections in the context of one-step imputation.

\protect\phantomsection\label{supplementary-material}
\bigskip

\begin{center}

{\large\bf SUPPLEMENTARY MATERIAL}

\end{center}

\appendix

\section{\texorpdfstring{Asymptotic properties of the plug-in estimator
of \(\E[U(X,A,Z;\thetalim)\mid
\Delta = 0, A = a]\)}{Asymptotic properties of the plug-in estimator of \textbackslash E{[}U(X,A,Z;\textbackslash thetalim)\textbackslash mid
\textbackslash Delta = 0, A = a{]}}}\label{sec-supp-asymp}

Recall that \(\PsiU = \E[U(X,A,Z;\thetalim)\mid \Delta = 0, A = a]\) and
let \begin{align*}
\widehat \PsiU = \frac{1}{n} \sum_{i=1}^n \frac{I(A_i=a)}{\widehat
g(a)}\frac{1-\Delta_i}{1 - \widehat
S(A_i)} U(X_i, A_i,Z_i; \widehat \theta)
\end{align*} denote the plug-in estimator. It follows from direct
calculation that we can make the following von Mises expansion
\begin{align}
\widehat \PsiU - \PsiU =& \mathbb{P}_n
\frac{I(A=a)}{g_0(a)}\frac{1-\Delta}{1-S_0(A)}U(X,A,Z; \thetalim) - \PsiU
\nonumber \\
+& \mathbb{P}\left[\frac{I(A=a)}{\widehat g(a)}\frac{1-\Delta}{1 - \widehat
S(A)} U(X, A,Z; \widehat \theta) - \PsiU\right] \label{eq:UpluginSecOrderRem} \\
+& (\mathbb{P}_n - \mathbb{P}) \left[\frac{I(A=a)}{\widehat g(a)}\frac{1-\Delta}{1 - \widehat
S(A)} U(X, A,Z; \widehat \theta) - \frac{I(A=a)}{g_0(a)}\frac{1-\Delta}{1 - S_0(A)} U(X,
A,Z; \thetalim) \right], \label{eq:UpluginEmRem}
\end{align} where \(\mathbb{P}_n\) denotes the empirical average over
the \(n\) iid observations, and \(\mathbb{P}\) denotes the integral over
the true probability distribution of \(X,A,Z\) ignoring the
stochasticity of the nuisance models, i.e., \(\widehat g, \widehat S\),
and \(\widehat \theta\). The empirical process remainder term
\eqref{eq:UpluginEmRem} can be shown to be \(o_{P_0}(n^{-1/2})\) under
mild regularity conditions (Donsker class conditions)
\citep{van1998asymptotic}. The second order remainder
\eqref{eq:UpluginSecOrderRem} equals \begin{align}
&\mathbb{P}\left[\frac{I(A=a)}{\widehat g(a)}\frac{1-\Delta}{1 - \widehat
S(A)} U(X, A,Z; \widehat \theta) - \frac{I(A=a)}{g_0(a)}\frac{1-\Delta}{1 - S_0(A)} U(X,
A,Z; \thetalim)\right] \nonumber \\
&= \mathbb{P}\left[\frac{I(A=a)}{g_0(a)}\frac{1-\Delta}{1 -
S_0(A)} \left\{U(X, A,Z; \widehat \theta) -  U(X,
A,Z; \thetalim)\right\}\right] \label{eq:UpluginSecRem1}\\
&+ \mathbb{P}\left[\left(\frac{I(A=a)}{\widehat g(a)}\frac{1-\Delta}{1 - \widehat
S(A)} - \frac{I(A=a)}{g_0(a)}\frac{1-\Delta}{1 - S_0(A)} \right)\left\{U(X,
A,Z; \widehat \theta) - U(X,A,Z; \thetalim) \right\}\right]  \label{eq:UpluginSecRem2}\\
&+\mathbb{P}\left[\left(\frac{I(A=a)}{\widehat g(a)}\frac{1-\Delta}{1 - \widehat
S(A)} - \frac{I(A=a)}{g_0(a)}\frac{1-\Delta}{1 - S_0(A)} \right) U(X,
A,Z; \thetalim)\right].  \label{eq:UpluginSecRem3}
\end{align} For \eqref{eq:UpluginSecRem1}, using the linear
decomposition of \(\widehat
\theta\) in \eqref{eq:UthetaIF}, we see that \begin{align*}
&\mathbb{P}\left[\frac{I(A=a)}{g_0(a)}\frac{1-\Delta}{1 -
S_0(A)} \left\{U(X, A,Z; \widehat \theta) -  U(X,
A,Z; \thetalim)\right\}\right] \\
= &\frac{1}{n}\sum_{i=1}^n \PP\left(\frac{I(A=a)}{g_0(a)}\frac{1-\Delta}{1 -
S_0(A)} \nabla_\theta U(X,A,Z; \thetalim)\right)\epsilon(X_i,A_i,Z_i,\Delta_i,
\Delta_i Y_i; P_0) + o_{P_0}(n^{-1/2}).
\end{align*} Considering \eqref{eq:UpluginSecRem2}, we first not that
\(\widehat g(a)>C\) and \(1-\widehat S(a)>C\) almost surely for some
constant \(C>0\). Thus, an application of the Cauchy-Schwarz inequality
yields that the term is \(o_{P_0}(n^{-1/2})\) because
\(\lVert \widehat g(a) - g_0(a)
\rVert_{2,P_0}\), \(\lVert \widehat S(a) - S_0(a)
\rVert_{2,P_0}\) are \(O_{P_0}(n^{-1/2})\), and
\(\lVert U(X,A,Z;\widehat \theta) - U(X,A,Z;\thetalim) \rVert_{2,P_0}\)
is \(o_{P_0}(1)\).

Finally, for \eqref{eq:UpluginSecRem3}, using Slutsky's theorem, we see
that \begin{align*}
&\mathbb{P}\left[\left(\frac{I(A=a)}{\widehat g(a)}\frac{1-\Delta}{1 - \widehat
S(A)} - \frac{I(A=a)}{g_0(a)}\frac{1-\Delta}{1 - S_0(A)} \right) U(X,
A,Z; \thetalim)\right]\\
= &\left(\frac{1}{\widehat g(a)}\frac{1}{1 - \widehat
S(a)} - \frac{1}{g_0(a)}\frac{1}{1 - S_0(a)}\right) \mathbb{P}\left[I(A=a)(1-\Delta) U(X,
A,Z; \thetalim)\right]\\
&= \left(\frac{g_0(a)(1-S_0(a)) - \widehat g(a) (1-\widehat S(a))}{\widehat g(a) g_0(a) (1 - \widehat
S(a))(1-S_0(a))}\right) \mathbb{P}\left[I(A=a)(1-\Delta) U(X,
A,Z; \thetalim)\right]\\
&= \frac{1}{n}\sum_{i = 1}^n \left(\frac{g_0(a)(1-S_0(a)) - I(A_i=a)(1-\Delta_i)}{g_0(a)
(1-S_0(a))}\right) \PsiU + o_{P_0}(n^{-1/2}).
\end{align*} Collecting all of the pieces we get that \begin{align*}
\sqrt{n}(\widehat \PsiU - \PsiU) =& n^{-1/2} \sum_{i = 1}^n \frac{I(A_i=a)}{g_0(a)}\frac{1-\Delta_i}{1-S_0(a)}\left\{U(X_i,A_i,Z_i, \thetalim)
-\PsiU \right\}\\
+& n^{-1/2} \sum_{i = 1}^n \PP\left(\frac{I(A=a)}{g_0(a)}\frac{1-\Delta}{1 -
S_0(a)}\nabla_\theta U(X,A,Z; \thetalim)\right)\epsilon(X_i,A_i,Z_i,\Delta_i,
\Delta_i Y_i; P_0)\\
&+ o_{P_0}(1).
\end{align*} Thus, the Central Limit Theorem provides valid asymptotic
inference for the plug-in estimator based on the above influence
function decomposition.

\section{\texorpdfstring{RCT-augmented one-step estimator of
\(\E[U(X,A,Z;\thetalim)\mid \Delta = 0, A = a]\)}{RCT-augmented one-step estimator of \textbackslash E{[}U(X,A,Z;\textbackslash thetalim)\textbackslash mid \textbackslash Delta = 0, A = a{]}}}\label{sec-supp-asymp-2}

Due to randomization, the baseline covariates \(X\) are independent of
the treatment indicator \(A\). This makes it possible to utilize
information on baseline covariates to increase efficiency when
estimating treatment effects. This is also the case for the target
parameter \(\PsiU =
\E[U(X,A,Z;\thetalim)\mid \Delta = 0, A = a]\).

Assuming that the imputation parameter \(\thetalim\) is known, the
influence function for the empirical mean estimator of
\(\PsiU = \E[U(X,A,Z;\thetalim)
\mid \Delta = 0, A = a]\) is \begin{align*}
\frac{I(A=a)}{g_0(a)}\frac{1-\Delta}{1-S_0(a)}\Big(U(X,A,Z;\thetalim)- \PsiU\Big).
\end{align*} A projection of the influence function onto the RCT tangent
space spanned by elements of the form \((A-g_0(1))h(X)\) \citep[Ch.
5]{tsiatis2006semiparametric} yields that the efficient influence
function is given by \begin{align*}
&\frac{I(A=a)}{g_0(a)}\frac{1-\Delta}{1-S_0(A)}\Big(U(X,A,Z;\thetalim)-
\PsiU\Big) \\
+ &\frac{g_0(a)-I(A=a)}{g_0(a)}\frac{1-S_0(X,a)}{1-S_0(a)}\Big(H_0(X,a;\thetalim)- \PsiU\Big),
\end{align*} where \begin{align*}
H_0(X,a;\thetalim) = \E\left[U(X,A,Z;\thetalim)\mid X, A=a\right].
\end{align*} Thus, we construct an RCT-augmented one-step estimator
\begin{align*}
\widetilde \PsiU = \widehat \PsiU + \mathbb{P}_n \left(\frac{\widehat
g(a)-I(A=a)}{\widehat g(a)}\frac{1-\widehat S(X,a)}{1-\widehat S(a)}\left(\widehat H(X,a)-
\widehat \PsiU\right)\right),
\end{align*} where \(\widehat \PsiU\) is the plug-in estimator given in
Section~\ref{sec-supp-asymp}. The second order remainder related to the
augmentation term of the RCT-augmented one-step estimator is
\begin{align*}
&\PP \left[\frac{\widehat
g(a)-I(A=a)}{\widehat g(a)}\frac{1-\widehat S(X,a)}{1-\widehat S(a)}\left\{\widehat H(X,a)-
\widehat \PsiU\right\} \right]\\
=& \frac{\widehat g(a) - g_0(a)}{g_0(a)} \PP \left[\frac{1- S^\ast(X,a)}{1-
S_0(a)} \left\{H^\ast(X,a)-\PsiU\right\}\right]\\
+& \frac{\widehat g(a) - g_0(a)}{g_0(a)} \PP \left[\frac{1-\widehat S(X,a)}{1-\widehat
S(a)}\left\{\widehat H(X,a) -\widehat \PsiU \right\} -\frac{1- S^\ast(X,a)}{1-
S_0(a)} \left\{H^\ast(X,a) -\PsiU \right\}\right]\\
=& \frac{\widehat g(a) - g_0(a)}{g_0(a)} \PP \left[\frac{1- S^\ast(X,a)}{1-
S_0(a)} \left\{H^\ast(X,a)-\PsiU\right\}\right]\\
+& o_{P_0}(n^{-1/2}),
\end{align*} where we make the weak assumption that
\(\lVert \widehat S(X,a) - S^\ast(X,a) \rVert_{2,P_0} =
o_{P_0}(1)\) and
\(\lVert \widehat H(X,a) - H^\ast(X,a) \rVert_{2,P_0} = o_{P_0}(1)\).
Thus, the influence function for the RCT-augmented one-step estimator is
\begin{align*}
& \frac{I(A=a)}{g_0(a)}\frac{1-\Delta}{1-S_0(A)}\left\{U(X,A,Z; \thetalim)
-\PsiU\right\}\\
+& \PP\left(\frac{I(A=a)}{g_0(a)}\frac{1-\Delta}{1 -
S_0(A)}\nabla_\theta U(X,A,Z; \thetalim)\right)\epsilon(X,A,\Delta,
\Delta Y; P_0)\\
+&\frac{g_0(a)-I(A=a)}{g_0(a)}\frac{1-S^\ast(X,a)}{1-S_0(a)}\left\{H^\ast(X,a;\thetalim)- \PsiU\right\}\\
+& \frac{I(A=a) - g_0(a)}{g_0(a)} \PP \left[\frac{1- S^\ast(X,a)}{1-
S_0(a)} \left\{H^\ast(X,a; \thetalim)-\PsiU\right\}\right].
\end{align*}

\section{Direct one-step imputation estimation}\label{sec-supp-direct}

Instead of estimating each of the sub-target parameters as presented in
Section~\ref{sec-estimation}, it is possible to directly estimate
\(\PsiTildeY = \E[\tilde Y(\thetalim) \mid A =
a]\) or even the contrast \(\E[\tilde Y(\thetalim) \mid A =
1] - \E[\tilde Y(\thetalim) \mid A = 0]\).

Consider the one-step estimator \begin{align*}
\widehat \PsiTildeY =
\frac{1}{n}\sum_{i = 1}^n \frac{I(A_i = a)}{\widehat g(a)} \left\{\Delta_i Y_i +
(1-\Delta_i)U(X_i, A_i, Z_i;\widehat \theta) -
\widehat Q(X_i, A_i) \right\} + \widehat Q(X_i, a).
\end{align*} Then, the same arguments as in Section~\ref{sec-supp-asymp}
and Section~\ref{sec-supp-asymp-2} give \begin{align*}
\sqrt{n}\left(\widehat \PsiTildeY - \PsiTildeY\right) = n^{-1/2} \sum_{i = 1}^n
\phiTildeY (X_i, A_i, Z_i, \Delta_i, \Delta_i Y_i; P^\ast) + o_{P_0}(1),
\end{align*} where \begin{align*}
\phiTildeY &(X, A, Z, \Delta, \Delta Y; P^\ast) \\
&= \frac{I(A = a)}{g_0(a)}
\left\{\Delta Y + (1-\Delta)U(X, A, Z;\thetalim) - Q^\ast(X, A) \right\} + Q^\ast(X, a) - \PsiTildeY \\
&\quad + \frac{g_0(a) - I(A=a)}{g_0(a)}
\E\left[\frac{I(A=a)}{g_0(a)}\left\{\Delta Y + (1-\Delta)U(X,A,Z;\thetalim) -
Q^\ast(X, A) \right\}\right]\\
&\quad + \E \left[\frac{I(A=a)}{g_0(a)}(1-\Delta)\nabla_\theta U(X,A,Z; \thetalim)\right]\epsilon(X,A,Z,\Delta,
\Delta Y; P_0).
\end{align*} If
\(Q^\ast(X, a) = \E[\Delta Y + (1-\Delta)U(X,A,Z;\thetalim) \mid X, A = a]\),
the estimator is asymptotically efficient for \(\PsiTildeY\) in the case
where \(\thetalim\) is known. When \(\thetalim\) is estimated, the third
term of the influence function contributes additional variance.

\section{Target parameter implied by multiple
imputation}\label{sec-supp-mi-target}

For simplicity, let \((Y_i, X_i)\) be iid data. Assume that the
conditional distribution of the outcome \(Y\) given \(X=x\) has density
\(f_0(y \mid x)\). Consider a parametric model \(f(y \mid x; \theta)\)
which is misspecified such that the estimator \(\widehat\theta\)
converges to \(\thetalim\) and
\(f_0 \neq f(\cdot \mid \cdot ; \thetalim)\). Let
\(U(X; \theta) = \int y f(y \mid X; \theta)\, dy\) denote the
\emph{model-implied} mean.

We consider estimation via (frequentist) multiple imputation (MI), with
\(\widehat\theta\) estimated with an M-estimator, e.g.~MLE. Let
\(U_{ij}\), \(j=1,\ldots,m\), be independent draws from the conditional
density \(f(y \mid X_i; \widehat\theta)\). We need to show that
\begin{align*} 
\frac{1}{nm}\sum_{i=1}^n
\sum_{j=1}^m U_{ij} - \E[U(X;\thetalim)] = o_P(1). 
\end{align*}

We decompose the difference as \begin{align*}
&\frac{1}{nm}\sum_{i=1}^n \sum_{j=1}^m U_{ij} -
\E[U(X;\thetalim)] \\
&= \underbrace{\frac{1}{n}\sum_{i=1}^n \frac{1}{m} \sum_{j=1}^m U_{ij} -
\frac{1}{n}\sum_{i=1}^n U(X_i;\widehat\theta_n)}_{(A)~\text{Monte-Carlo error}} \\
&+
\underbrace{\frac{1}{n}\sum_{i=1}^n \left\{U(X_i; \widehat\theta_n) - U(X_i;
\thetalim)\right\}}_{(B)~\text{Parameter estimation error}} \\
&+ \underbrace{\frac{1}{n}\sum_{i=1}^n U(X_i; \thetalim) - \E[U(X;
\thetalim)]}_{(C)~\text{Sampling error}}.
\end{align*}

The term \((C)\) is immediately \(o_P(1)\) by the law of large numbers.
To control the term \((B)\) we assume a local Lipschitz condition, in
the sense that there exists a neighborhood \(K(\thetalim)\) and an
integrable envelope function \(L(X)\), \(\E[L(X)] < \infty\), such that
\begin{align*}
\sup_{\theta\in K(\thetalim)}\lVert\nabla_\theta U(x;\theta)\rVert \leq L(x)
\end{align*} almost surely. Here \(\lVert\cdot\rVert\) denotes the usual
Euclidean norm. From the Mean Value Theorem we have that \begin{align*}
U(X; \widehat\theta_n) - U(X;\thetalim) =
\nabla_\theta U(X; \overline{\theta})^\top(\widehat\theta_n - \thetalim),
\end{align*} where \(\overline{\theta}\) lies on the line segment
between \(\widehat{\theta}_n\) and \(\thetalim\). It then follows that
on the set \(K(\thetalim)\) \begin{align*}
\lvert(B)\rvert \leq \lVert\widehat{\theta}_n-\thetalim\rVert
\frac{1}{n}\sum_{i=1}^n\lVert\nabla_\theta U(X_i; \overline{\theta}_i)\rVert
\leq
\lVert\widehat{\theta}_n-\thetalim\rVert
\frac{1}{n}\sum_{i=1}^n L(X_i),
\end{align*} where we have used the local Lipschitz envelope condition.
The term \(n^{-1}\sum L(X_i)\) converges in probability by the law of
large numbers to \(\E[L(X)] < \infty\), and
\(\lVert\widehat{\theta}_n-\thetalim\rVert = O_P(n^{-1/2})\); hence the
right-hand side is \(o_P(1)\). It therefore follows that
\((B) = o_P(1)\), since \begin{align*}
P(\lvert (B)\rvert > \varepsilon) \leq P(\lvert (B)\rvert > \varepsilon, \widehat\theta \in K(\thetalim)) +
P(\widehat\theta \notin K(\thetalim)).
\end{align*}

For the Monte-Carlo error term, let \begin{align*}
\epsilon_i = \frac{1}{m}\sum_{j=1}^m U_{ij} - U(X_i;\widehat\theta_n),
\end{align*} so that \((A) = n^{-1}\sum_{i=1}^n \epsilon_i\). Let
\(\mathcal{F}_n := \sigma\!\left(\{(X_i)\}_{i=1}^n,\widehat\theta_n\right)\).
The key is now that conditionally on \(\mathcal{F}_n\), the imputations
\(U_{ij}\sim f(\cdot\mid
X_i;\widehat\theta_n)\) are independent across \(i\) and \(j\), so the
residuals \(\epsilon_i\) are conditionally independent with
\begin{align*}
\E[\epsilon_i\mid \mathcal F_n] = 0,
\qquad
\var(\epsilon_i\mid \mathcal{F}_n) = \frac{v(X_i;\widehat\theta_n)}{m},
\end{align*} where \(v(x;\theta) := \var(Y\mid X=x;\theta)\) is the
/model-implied/ variance. Hence \begin{align*}
\E\!\left[(A)^2\,\big|\,\mathcal{F}_n\right]
= \frac{1}{n^2}\sum_{i=1}^n \var(\epsilon_i\mid\mathcal{F}_n)
= \frac{1}{n^2 m}\sum_{i=1}^n v(X_i;\widehat\theta_n).
\end{align*} Assume the existence of a variance envelope, \(V(x)\), such
that \begin{align*}
\sup_{\theta\in K(\thetalim)} v(x;\theta) \le V(x),\qquad \E[V(X)]<\infty.
\end{align*} On the event \(\{\widehat\theta_n\in K(\thetalim)\}\) we
have \(v(X_i;\widehat\theta_n)\le V(X_i)\) for every \(i\). Therefore,
on this event, \begin{align*}
\E\!\left[(A)^2\,\big|\,\mathcal{F}_n\right] = \frac{1}{nm}\cdot O_P(1) = O_P\left((nm)^{-1}\right).
\end{align*} The conditional Chebyshev's inequality yields, for any
\(\epsilon>0\), \begin{align*}
P\left(|(A)|>\epsilon\, \big|\, \mathcal{F}_n\right) \le
\frac{\E[(A)^2\mid\mathcal F_n]}{\epsilon^2}
\end{align*} hence by the tower property \begin{align*}
P\left(|(A)| > \epsilon\right) \leq \frac{\E[V(X)]}{nm\epsilon^2} \to 0 \text{ as } nm\to\infty.
\end{align*} Finally, similarly as before for \((B)\), we note that
\begin{align*}
P(\lvert (A)\rvert > \varepsilon) \leq P(\lvert (A)\rvert > \varepsilon, \widehat\theta \in K(\thetalim)) +
P(\widehat\theta \notin K(\thetalim))
\end{align*} which then together with
\(P(\widehat\theta_n\notin K(\thetalim))\to 0\), gives \((A)=o_P(1)\) as
\(nm\to\infty\) (in fact, this also holds for \(m\) fixed as
\(n\to\infty\)).

As an important special case we consider the case where \(Y_i\)
conditionally on \(X_i\) follows an exponential family distribution with
density \begin{align*}
f(y|x; \theta, \phi) = \exp\left[
    \left\{y\eta(x; \theta)-b[\eta(x; \theta)]
    \right\}a(\phi)+c(y,\phi)
\right] 
\end{align*} i.e., with score with respect to \(\eta\) given by
\(\nabla_\eta \ell(\eta) = \{y-b'(\eta)\}/a(\phi)\), and it follows that
the conditional mean is given by \begin{align*}
\E[Y\mid X=x; \theta] = U(x; \theta) = b'(\eta(x; \theta)).
\end{align*}

For a generalized linear model with canonical link functions, \(g\), we
have \(\eta := \eta(x; \theta) = g(U\{x; \theta\}) = x^\top\theta\), and
\begin{align*}
\nabla_\theta U(x; \theta) = b''(x^\top\theta)x.
\end{align*} For example, for a Gaussian linear model we would have
\(g(x) = x\) and \(b(\eta) = \eta^2/2\), and thus
\(\nabla_\theta U(x; \theta) = x\), and it follows that the local
Lipschitz condition holds as long as \(\E\lVert X\rVert<\infty\), which
is implied by a finite-second-moment condition on \(X\) which is already
assumed for the \(\sqrt n\)-consistency of the M-estimator. The variance
envelope condition is automatically satisfied under variance homogeneity
\(v(x;\theta)=\sigma^2\). Similarly, for a logistic regression we have
\(g(x) = \log(x)-\log(1-x)\) and \(b(\eta) = \log(1+\exp(\eta))\), and
thus \(\nabla_\theta U(x; \theta) =
b''(x^\top\theta)\,x = \exp(x^\top \theta)\{1+\exp(x^\top\theta)\}^{-2}\,x\)
which again obeys the Lipschitz condition as long as
\(\E\lVert X\rVert<\infty\), and we can choose the variance envelope
\(V(x)=\tfrac{1}{4}\).

\section{Full data-generating mechanism for the retrieved-dropout
simulation}\label{sec-supp-siminter-dgp}

This section states the full parametrization of the data-generating
mechanism used in the retrieved-dropout simulation example in
Section~\ref{sec-simdrop}.

\textbf{Baseline covariates.} \begin{align*}
\pi_{X_1} &= 0.2, \\
\mu_{X_2}(X_1) &= 0.15 - 0.08 X_1, \\
\mu_{X_3}(X_1, X_2) &= 0.3 X_1 - 0.08 X_2 + 0.05 X_1 X_2.
\end{align*}

\textbf{Treatment discontinuation (\(\pi_{Z_1}\)).} The \(8\)-term
logistic regression is \begin{align*}
\mbox{logit}\, \pi_{Z_1}(A, X_1, X_2, X_3)
&= -1.4 - 0.1 A - 0.4 X_1 + 0.15 X_2 - 0.12 X_3 \\
&\quad + 0.08 A X_1 - 0.05 A X_2 - 0.15 A X_3.
\end{align*}

\textbf{Non-terminal event (\(\pi_{Z_2}\)).} The \(16\)-term logistic
regression is \begin{align*}
\mbox{logit}\, \pi_{Z_2}(A, X_1, X_2, X_3, Z_1)
&= 3.7 + 0.8 A - 4 Z_1 + 0.7 X_1 - 1.5 X_2 + 0.5 X_3 \\
&\quad - 0.6 A Z_1 - 0.9 A X_1 + 0.15 A X_2 + 0.05 A X_3 \\
&\quad - 0.35 Z_1 X_1 + 1.4 Z_1 X_2 - 0.4 Z_1 X_3 \\
&\quad + 0.7 A Z_1 X_1 - 0.25 A Z_1 X_2 - 0.1 A Z_1 X_3.
\end{align*}

\textbf{Missingness (\(\pi_{\text{mis}}\)).} The \(16\)-term logistic
regression is \begin{align*}
\mbox{logit}\, \pi_{\text{mis}}(A, X_1, X_2, X_3, Z_1)
&= -3.3 + 0 \cdot A + 4 Z_1 - 0.2 X_1 + 0.4 X_2 - 0.1 X_3 \\
&\quad - 0.25 A Z_1 + 0.04 A X_1 - 0.2 A X_2 + 0.3 A X_3 \\
&\quad - 0.5 Z_1 X_1 - 0.35 Z_1 X_2 + 0.1 Z_1 X_3 \\
&\quad + 0 \cdot A Z_1 X_1 + 0.2 A Z_1 X_2 - 0.3 A Z_1 X_3.
\end{align*}

\textbf{Outcome mean (\(\mu_Y\)).} The \(16\)-term linear function is
\begin{align*}
\mu_Y(A, X_1, X_2, X_3, Z_1)
&= 0.13 + 0.16 A - 0.17 Z_1 + 0.06 X_1 - 0.15 X_2 + 0.6 X_3 \\
&\quad - 0.025 A Z_1 + 0.04 A X_1 + 0.02 A X_2 - 0.06 A X_3 \\
&\quad - 0.15 Z_1 X_1 - 0.018 Z_1 X_2 + 0 \cdot Z_1 X_3 \\
&\quad + 0.45 A Z_1 X_1 - 0.1 A Z_1 X_2 + 0.1 A Z_1 X_3.
\end{align*}

\bibliography{bibliography}

\end{document}